\documentclass{ws-ijmpe}

\begin{document}

\markboth{Authors' Names}{Instructions for  
Typing Manuscripts (Paper's Title)}

\catchline{}{}{}{}{}
%
%%%%%%%%%%%%%%%%%%%%%%%%%%%%%%%%%%%%%%%%%%%%%%%%%%%%%%%%%%%%%%%%%%%%

\title{MICROSCOPIC DETERMINATIONS OF FISSION BARRIERS \\(MEAN-FIELD AND BEYOND) }

\author{\footnotesize A. DOBROWOLSKI, H. GOUTTE, J.-F. BERGER \footnote{CEA/DAM, Ile de France, DPTA,
Service de Physique nucl\`eaire, Bruy\`eres-Le-Ch\^atel, France}}

\address{CEA/DAM, Ile de France, DPTA/Service de Physique Nucl\'eaire\\
BP 12, 91680 Bruy\`eres-Le-Ch\^atel, France}

\maketitle

\begin{history}
\received{(received date)}
\revised{(revised date)}
%\accepted{(Day Month Year)}
%\comby{(xxxxxxxxxx)}
\end{history}

\begin{abstract}

With a help of the selfconsistent Hartree-Fock-Bogoliubov (HFB) approach with the D1S effective Gogny interaction and 
the Generator 
Coordinate Method (GCM) we incorporate the transverse collective vibrations to the one-dimensional model 
of the fission-barrier penetrability based on the traditional WKB method. The average fission barrier
corresponding to the least-energy path in the two-dimensional potential energy landscape as function
of quadrupole and octupole degrees of freedom is modified by the influence of the transverse collective vibrations
along the nuclear path to fission. The set of transverse vibrational states built in the fission 
valley corresponding to a successively increasing nuclear elongation produces the new energy barrier which is 
compared with the least-energy barrier. 
These collective states are given as the eigensolutions of the GCM purely vibrational 
Hamiltonian. In addition, the influence of the collective inertia on the fission properties is displayed, 
and it turns out to be the decisive condition for the possible transitions between different fission valleys. 

\end{abstract}

\section{Introduction}

The phenomenon of nuclear fission is a complex process which involves the evolution of a nucleus over the multidimensional
potential energy barrier emerged as a result of the competition between a short range attractive nucleon-nucleon interaction 
and the long range repulsive Coulomb forces. 
The problem of determining the most probable one-dimensional effective nuclear path to fission as an essential 
ingredient of the traditional WKB calculations of fission half-lives has been the subject of scientific efforts in 
low-energy nuclear physics since the time of the discovery of the nuclear fission phenomenon.

It is clear that the commonly used WKB approach to estimate the fission barrier penetrability 
gives a discrepancy of fission half-lives up to a few orders of magnitude compared to experimental data.
This fact points out on some essential deficiency of this approach.
Above all, one should realize that the WKB approximation, based on the assumption that the nuclear wave function
evolves, starting from the ground-state or isomeric minimum, along a one-dimensional (least-action or least-energy) 
path 
toward the scission configuration in the multidimensional potential energy 
landscape, neglects the influence of transverse collective vibrations of the nuclear surface. 
These vibrations which are, in general, of different multipolarity, amplitude and frequency, depending on the 
actual nuclear configuration, always go along with the elongation of the nucleus on its way to fission due 
to the specific valley-like shape of the potential along the fission path.  
It is then obvious that the total collective energy is partitioned between all collective modes 
occurring along the nuclear path to fission. Once the transverse collective modes are present, the  
effective fission barrier has to be understood as being determined by the set of
the lowest vibrational states, built in the one-dimensional collective-potential well
taken, for a given elongation, as the transverse cross section through the fission valley. 
 
A more fundamental approach would consist in assuming that the probability current related to the 
time-dependent nuclear wave function propagating essentially along the one-dimensional fission path has also 
non-zero components in all possible transverse directions.

The above mentioned effect, by its influence on the fission-barrier shape, is then expected to 
modify the estimates of fission half-lives within the traditional WKB method in a substantial way. 

In the studies presented below we restrict ourselves to discuss the effect of transverse vibrations only along the 
least-energy asymmetric paths (with a substantial mass asymmetry of the fission fragments) which corresponds 
to the minimal potential energy as function of the elongation.
This particular approach ensures that all vibrational states of the transverse mode possess higher energy
than the corresponding collective-potential minimum.  

For a reasonable treatment of the collective properties of fissioning actinide nuclei the minimal set of 
collective variables has to comprise, at least, the nuclear elongation, left--right asymmetry and possibly 
non-axiality. 
In the present work we study the least-energy paths, least-energy with transverse vibrations path and the 
least-action fission 
path for two actinide nuclei $^{226}$Th and $^{238}$U which in their ground states are practically stable 
against spontaneous fission. Already their shape-isomeric states can decay through spontaneous 
fission, a process which is competitive to the $\gamma-$back decay as shown in Ref.~\cite{BLCh06}. 
Since for those nuclei non-axial shapes are mostly important for configurations 
in the vicinity of the inner saddle point, (see e.g. Ref.~\cite{KPADJB}), the choice of only the elongation and the 
left-right asymmetry deformation degrees of freedom to uniquely describe nuclear states beyond the isomeric minimum 
seems to be entirely justified. 

In order to take into account the collective modes the configurations mixing of pure Hartree-Fock-Bogoliubov (HFB) states 
in the total collective wave function is assumed. Applying the Generator Coordinate Method (GCM) together with the 
Gaussian Overlap Approximation (GOA) one determines the vibrational collective states by diagonalizing 
the GCM collective Hamiltonian. The knowledge of eigensolutions of that Hamiltonian is essential for the problem of
the transverse vibrations we are going to discuss in this paper. 

%%%%%%%%%%%%%%%%%%%%%%%%%%%%%%%%%%%%%%%%%%%%%%%%%%%%%%%%%%%%%%%%%%%%%%%%%%%%%%%%

\section{The Hartree-Fock-Bogoliubov method}
\label{HFBapp}
The potential energy landscapes and nuclear intrinsic states of the fissioning nucleus 
are determined in constrained HFB calculations with the D1S Gogny force \cite{Gogny}
where the axially deformed nuclear states are described by quadrupole and octupole moments $\{q_2,q_3\}$ 
which create respectively the elongation and mass asymmetry of a nucleus.
These intrinsic states $|\Psi_q\rangle$ are defined in terms of quasi-particle vacuum states of the form 
(see e.g. Ref.~\cite{Ber-Gir})
\begin{eqnarray}
&&|\Psi_q\rangle=\prod\limits_i^{} \eta_i^+\,|0\rangle,\nonumber\\
&&\eta_i^+\,|\Psi_q\rangle=0,
\label{qp_func}\end{eqnarray}
where the quasi-particle creation operators $\eta_i^+$ are defined via the general Bogoliubov transformation.
The explicit form of the operators $\eta_i^+$ can be deduced from the variational principle
\begin{equation}
\delta\langle  \Psi(q_{2},q_{3})|\hat H-\lambda_N\hat N-\lambda_Z\hat Z-
\sum_i\lambda_i\hat Q_i|\Psi(q_{2},q_{3}) \rangle=0,
\label{variat}\end{equation}
where $\hat H$ is the many-body Hamiltonian built with the finite range 
effective Gogny force. The Lagrange multipliers $\lambda_N$, $\lambda_Z$ associated with the constraints 
on the neutron and proton number and the average nuclear deformations $q_i$ are calculated 
directly from the relations:
\begin{eqnarray}
&&\langle\Psi(q_{2},q_{3})|\hat N|\Psi(q_{2},q_{3})\rangle = N ,\nonumber\\
&&\langle\Psi(q_{2},q_{3})|\hat Z|\Psi(q_{2},q_{3})\rangle = Z ,\nonumber\\
&&\langle\Psi(q_{2},q_{3})|\hat Q_{i0}|\Psi(q_{2},q_{3})\rangle = q_i ,\nonumber\\
&&\langle\hat Q_{10}\rangle=\sum\limits_{i=1}^{N+Z} \langle z_i \rangle = 0 ,
\label{def_op}\end{eqnarray}
where the last condition ensures that the center of mass of the system is fixed at the origin of the 
coordinate system. 
The multipole operators $\hat Q_{20}$ and $\hat Q_{30}$ in eq.~(\ref{def_op}) are defined as 
\begin{eqnarray}
&&\hat Q_{20}=\sqrt{\frac{16\pi}{5}}\sum\limits_{i=1}^{N+Z} r_i^2\hat Y_{20}=\sum\limits_{i=1}^{N+Z}(2z_i^2-x_i^2-y_i^2),
\nonumber\\
&&\hat Q_{30}=\sqrt{\frac{4\pi}{7}}\sum\limits_{i=1}^{N+Z} r_i^3\hat Y_{30}=\sum\limits_{i=1}^{N+Z}\bigg[z_i^3-
\frac{3}{2}z_i(x_i^2+y_i^2)\bigg].
\end{eqnarray}

Equation (\ref{variat}) leads to the usual HFB equations which are solved by iterative diagonalization
of the HFB Hamiltonian matrix expressed in the axial harmonic-oscillator double-centered basis until the 
required convergence is achieved.
\begin{figure}[th]
\centerline{\psfig{file=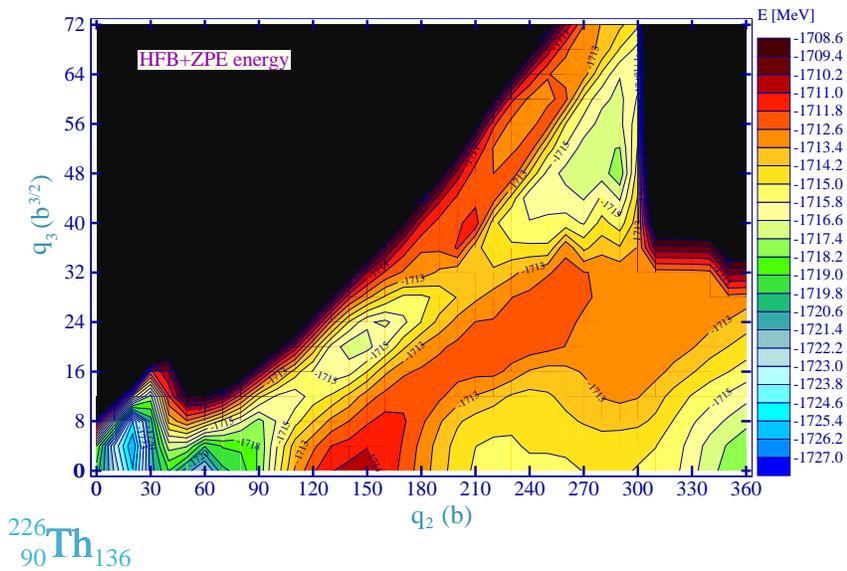,angle=270,width=11.cm}}
%\hspace{-3mm}\includegraphics[width=2.4cm, angle=270]{map_226.ps}
%\includegraphics[width=2.4cm, angle=270]{map_238.ps}
\caption{Collective (corrected by the zero-point movement, see text) potential-energy landscape for the 
actinide nucleus $^{226}$Th as function of quadrupole $q_2$ and octupole $q_3$ deformations.}
\label{maps1}\end{figure}
\begin{figure}[th]
\centerline{\psfig{file=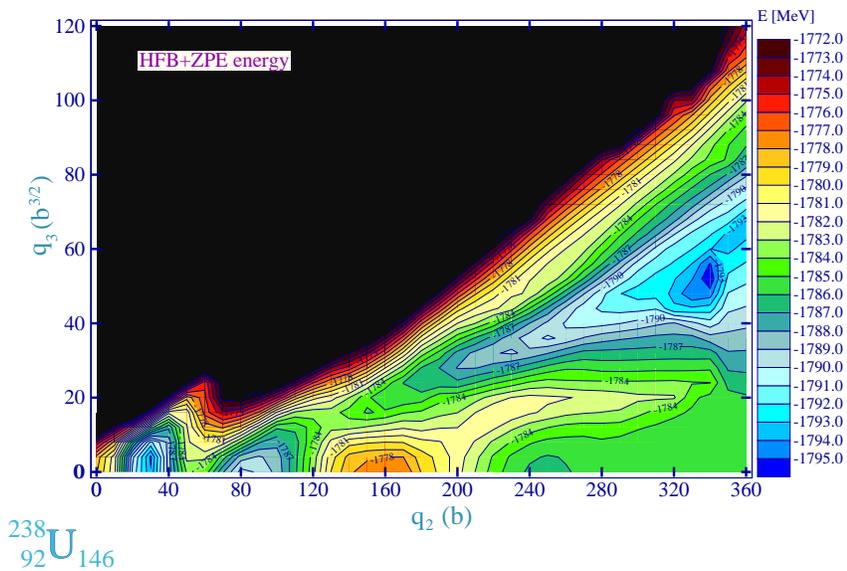, angle=270,width=11.cm}}
\vspace*{8pt}
\caption{Same as in Fig.~\ref{maps1} for the $^{238}$U nucleus.}
\label{maps2}\end{figure}
%\vspace{-3mm}\hspace{-0.5cm}\begin{minipage}[l]{1.025\textwidth}
%
Considering the symmetry properties of the HFB Hamiltonian one concludes that the entire potential-energy landscape 
is reflection symmetric with respect to $q_3=0$. This directly implies that in both cases, $^{226}$Th and $^{238}$U, 
presented in Figs.~\ref{maps1} and \ref{maps2} one can immediately distinguish the symmetric and asymmetric potential energy valleys 
separated by the energy barrier appearing for well-elongated shapes. Both those valleys can be potentially treated as the 
nuclear fission paths leading either to the symmetric or (much more probable) asymmetric mass fragmentation 
(see Ref.~\cite{Goutte} for $^{238}$U nucleus). 
 
In a purely static approach the criterion which determines the fission path is the potential energy value 
which has to be minimized over all deformation degrees of freedom entering  
eq.~(\ref{variat}) except the leading parameter $q_2$.  
 Nevertheless, this static approach with the use of $\{q_2,q_3\}$ as the deformation parameters can lead to artificial 
transitions between different possible paths if one is only interested in the absolute minimum of the potential 
energy with respect to $q_3$.
Clearly, $q_3$ can vary only within the limited interval describing the pre-scission configurations shown in 
Figs.~\ref{maps1}, \ref{maps2} with iso-energy lines. 
The absolute energy minima can occasionally belong to the symmetric valley, unlike the majority of elongated states
as easily seen for several configurations of $^{226}$Th (i.e. from $q_2\approx 190\,b$ to $q_2\approx 230\,b$).
Such sudden transitions between different fission valleys, which turn out to be highly probable in several Fm isotopes,
strongly depend also on the collective inertia of the system and properties of the energy barrier between them. 
More reliable studies of such transitions can be performed with the help of the well known least-action formalism 
to be briefly presented in section \ref{LAP}.
%\end{minipage}

%%%%%%%%%%%%%%%%%%%%%%%%%%%%%%%%%%%%%%%%%%%%%%%%%%%%%%%%%%%%%%%%%%%%%%%%%%%

%                   BEYOND  THE  MEAN_FIELD  APPROACH

%%%%%%%%%%%%%%%%%%%%%%%%%%%%%%%%%%%%%%%%%%%%%%%%%%%%%%%%%%%%%%%%%%%%%%%%%%%
%
\section{Beyond the mean-field approach}

We assume that the nuclear states are superpositions of the constrained HFB wave functions
in the form given by the GCM theory (for review see Ref.~\cite{Ring-Schuck})
\begin{equation}
|\Phi(q_{2},q_{3})_i\rangle=\int \!\! \int dq_{2}\,dq_{3}\,f_i(q_{2},q_{3})\,|\Psi(q_{2},q_{3})\rangle ,
\label{mix}\end{equation}
where $\{q_{2},q_{3}\}$ are the set of the collective variables identified as the quadrupole 
and octupole deformations respectively. 
The variational principle with a many-body Hamiltonian applied to the wave function defined 
by eq.~(\ref{mix}) leads to the Hill-Wheeler equation (see Ref.~\cite{Hill-Wheeller})
\begin{equation}
\int \!\! \int dq'_2dq'_3 \big[ H(q_2,q_3,q'_2,q'_3)-E_i\,I(q_2,q_3,q'_2,q'_3)\big]\,f_i(q'_2,q'_3)=0
\label{GCMeq}\end{equation}
which yields the weight functions $f_i(q_{2},q_{3})$. 
Quantities entering eq.~(\ref{GCMeq}) are given as
\begin{eqnarray}
&&H(q_2,q_3,q'_2,q'_3)=\langle \Psi(q_{2},q_{3})|\hat H|\Psi(q'_{2},q'_{3})\rangle,\nonumber\\  
&&I(q_2,q_3,q'_2,q'_3)=\langle \Psi(q_{2},q_{3})|\Psi(q'_{2},q'_{3})\rangle
\label{overlapy}\end{eqnarray}
and denote respectively the GCM Hamiltonian and the norm kernel while $E_i$ is the energy associated 
with the state $|\Phi(q_{2},q_{3})_i\rangle$.
Instead of solving the integral equation (\ref{GCMeq}) one can transform it to a much simpler form
admitting the GOA approximation, (e.g. Ref.~\cite{GOA}). With this assumption eq.~(\ref{GCMeq}) reads
\begin{equation}
\hat{\mathcal H}_{coll}\,g_i(q_{2},q_{3}) = E_{coll}\,g_i(q_{2},q_{3}),
\end{equation}
where now $g_i(q_{2},q_{3})$ being the Gauss transform of $f_i(q_{2},q_{3})$ (see e.g. Ref.~\cite{Gozdz})
is the collective wave function while $\hat{\mathcal H}_{coll}$ is the collective Hamiltonian 
\begin{equation}
\hat{\mathcal H}_{coll}=-\frac{\hbar^2}{2}\sum\limits_{k,l=2}^3 \frac{\partial}{\partial q_{k}}
\frac{1}{B_{kl}(q_{2},q_{3})}\frac{\partial}{\partial q_{l}} + V_{{\rm HFB}}(q_{2},q_{3}) -
\sum\limits_{k,l=2}^3\Delta V_{kl}(q_{2},q_{3})
\label{hcoll}\end{equation}
which can be proved, in a straightforward calculations, to be formally identical to the collective Bohr Hamiltonian.
The quantities $V_{{\rm HFB}}(q_{2},q_{3})$, $\Delta V_{kl}(q_{2},q_{3})$, $B_{kl}$ are respectively
the constrained HFB deformation energy, the so-called zero-point (ZPE) vibrational and rotational energy corrections 
(see Refs.~\cite{ZPE,ZPE1}) and the inertia tensor associated with the quadrupole and octupole modes. 
The latter is calculated in the Inglis cranking model \cite{Inglis}.
The ZPE vibrational correction origins directly from the fact that 
overlaps $I$, eq.~(\ref{overlapy}) are not products of $\delta(q_2-q_2')\delta(q_3-q_3')$. 
The rotational ZPE correction, in turn, is connected to the restoration on average of the rotational symmetry in the 
collective wave function $|\Phi(q_{2},q_{3})_i\rangle$.
However, for the simplicity of the calculations the latter can be determined in an approximate manner as
\begin{equation}
\Delta V_{rot}=\sum\limits_{n=1}^3\bigg\langle \Phi(q_{2},q_{3})_i\bigg| \frac{\hat J_n^2}{2{\mathcal I_n}}
\bigg|\Phi(q_{2},q_{3})_i\bigg\rangle,
\end{equation}
where $\hat J_n$ and ${\mathcal I_n}$ denotes respectively the $n^{th}-$component of the total angular momentum
and the nuclear moment of inertia with respect to the $n^{th}$ axis. 

The shortly presented GCM+GOA procedure goes, obviously, beyond the standard mean-field HFB approach 
providing in return, by the assumption of the configuration mixing, the average treatment of 
the long-amplitude nucleon-nucleon correlations which are not taken into account within the mean-field theory. 

%%%%%%%%%%%%%%%%%%%%%%%%%%%%%%%%%%%%%%%%%%%%%%%%%%%%%%%%%%%%%%%%%%%%%%%%%%%

%                   LEAST  ACTION  APPROACH

%%%%%%%%%%%%%%%%%%%%%%%%%%%%%%%%%%%%%%%%%%%%%%%%%%%%%%%%%%%%%%%%%%%%%%%%%%%

\section{Least-action approach}
\label{LAP}

The static approach presented in section \ref{HFBapp} allowing for determining in an approximate way the
nuclear path to fission is not entirely realistic since it neglects the collective inertia properties of the 
system. It was proved that the local behavior of the mass tensor can significantly change the
fission path as compared to the one determined by the energy-minimization condition.
It is a well known fact (see e.g. Ref.~\cite{BPA361}) that, in case the collective inertia is taken into account, 
the most favorable transition between two neighbouring configurations corresponding to slightly different elongations passes
through states for which either the energy or the inertia tensor decreases more rapidly.

Let us define the action in the two-dimensional potential energy landscape along the trajectory $L_i$ 
as e.g. in Refs.~\cite{BPA361,actionInt} 
\begin{equation}
S(L_i)=2\int\limits_{s_1}^{s_2}\sqrt{2[V(s)-E_{coll}]B(s)}ds,
\label{action}\end{equation}
where $s=s(q_2,q_3)$ describes parametrically a trajectory $L_i$ in the $\{q_2,q_3\}$ deformation space. 
The points $s_1$ and $s_2$ correspond to the turning points of this trajectory for which $V(s)=E_{coll}$.
Within the GCM+GOA approach applied here the HFB potential energy is deepened by the zero-point vibrational and 
rotational corrections $\Delta V_{vib}$, $\Delta V_{rot}$ respectively
\begin{equation}
V \equiv V_{HFB}-\Delta V_{rot}-\Delta V_{vib}.
\label{coll-pot}\end{equation}
It should be noticed that as far as the vibrational ZPE correction is concerned, it changes only insignificantly 
the shape of the fission barrier by 
shifting it down in energy by approximately the same energy for all minimum-energy configurations along the 
fission path, whereas the rotational correction lowers the barrier height (approximately) proportionally to the elongation.
In the present calculations, it is assumed that the penetration of the fission barrier starts from the lowest
vibrational state $E_{coll}$ built in the isomeric potential-energy well and associated with the mode along the 
fission path. 
The {\it projection} of the inertia tensor $B_{ij}(q_2,q_3)$ on the trajectory $L_i$ is given in the standard way as
\begin{equation}
B(s)=\sum_{ij=2,3} B_{ij}(q_2,q_3)\frac{dq_i}{ds}\frac{dq_j}{ds}.
\label{mass-trans}\end{equation}
The determination of the fission path corresponding to a more realistic treatment when both the energy and the 
inertia parameters are treated on the same footing would consist in finding the trajectory $L_{min}$
which minimizes the action integral of eq.~(\ref{action}). 

\begin{figure}[th]
\hspace{-8mm}\begin{minipage}[c]{0.5\textwidth}
{\psfig{file=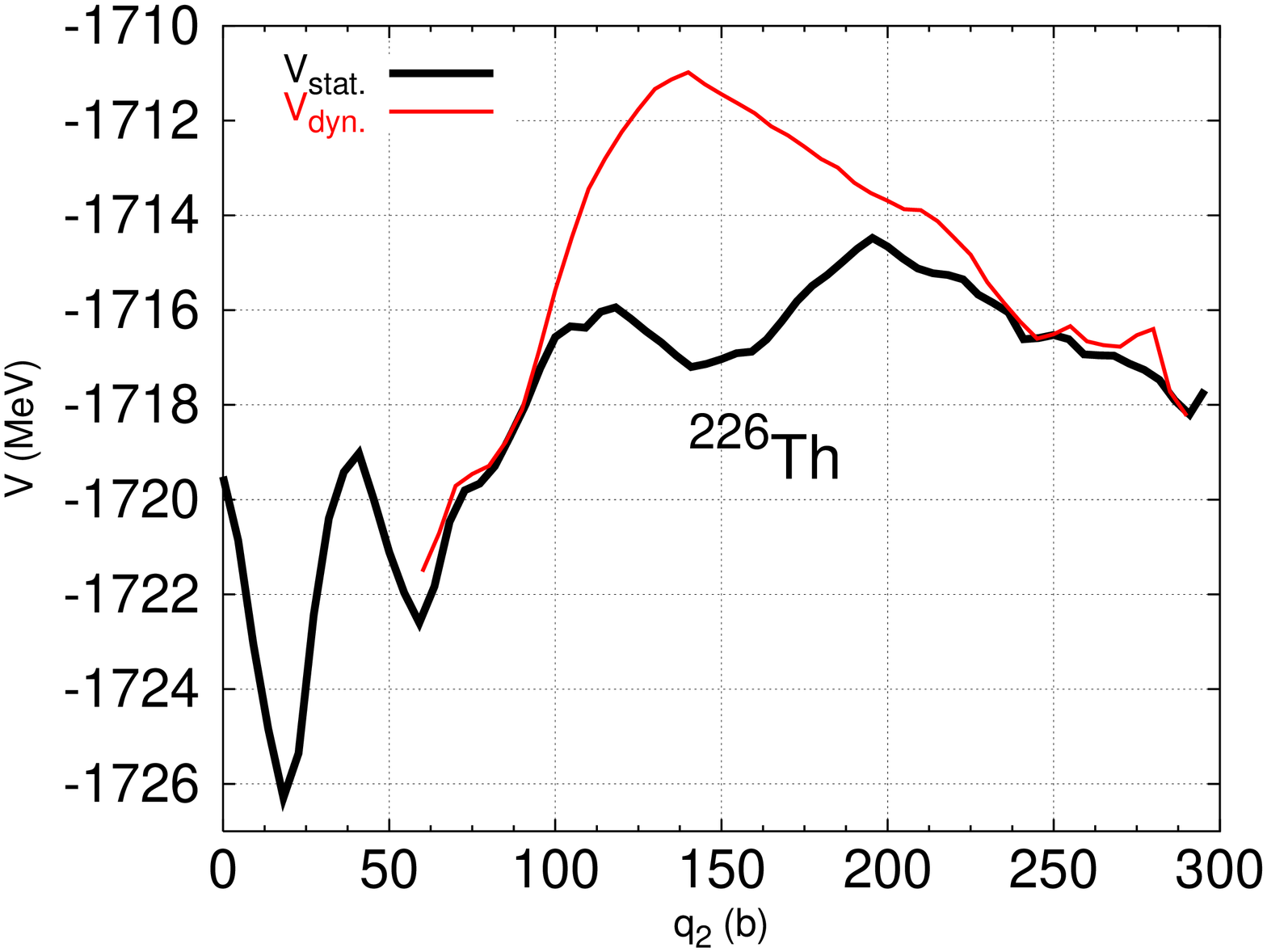, angle=270,width=7cm}}
%\hspace{-5mm}\includegraphics[width=4.5cm, angle=270]{porow226.eps}
\end{minipage}%
\begin{minipage}[c]{0.5\textwidth}
{\psfig{file=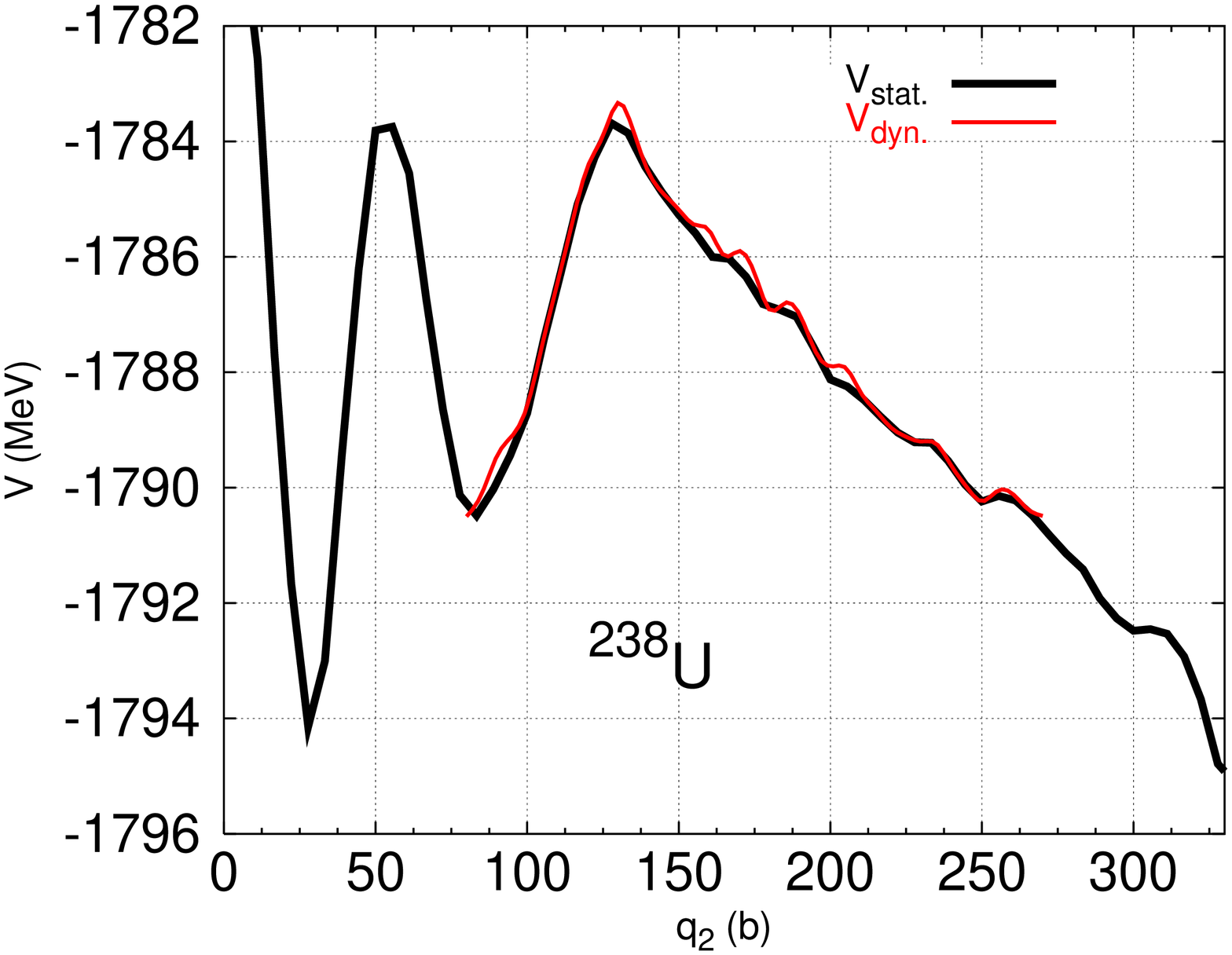, angle=270,width=7cm}}
%\hspace{-5mm}\includegraphics[width=4.5cm, angle=270]{porow238.eps}
\end{minipage} 
\vspace*{8pt}
\caption  {Potential energy barrier along the $L_{min}$ trajectory (thin) 
and the barrier corresponding to the least-energy trajectory (thick) for the 
actinide nuclei $^{226}$Th and $^{238}$U. }
\label{bar-dyna}\end{figure}
\begin{figure}[th]
\begin{minipage}[c]{0.5\textwidth}
\centerline{\psfig{file=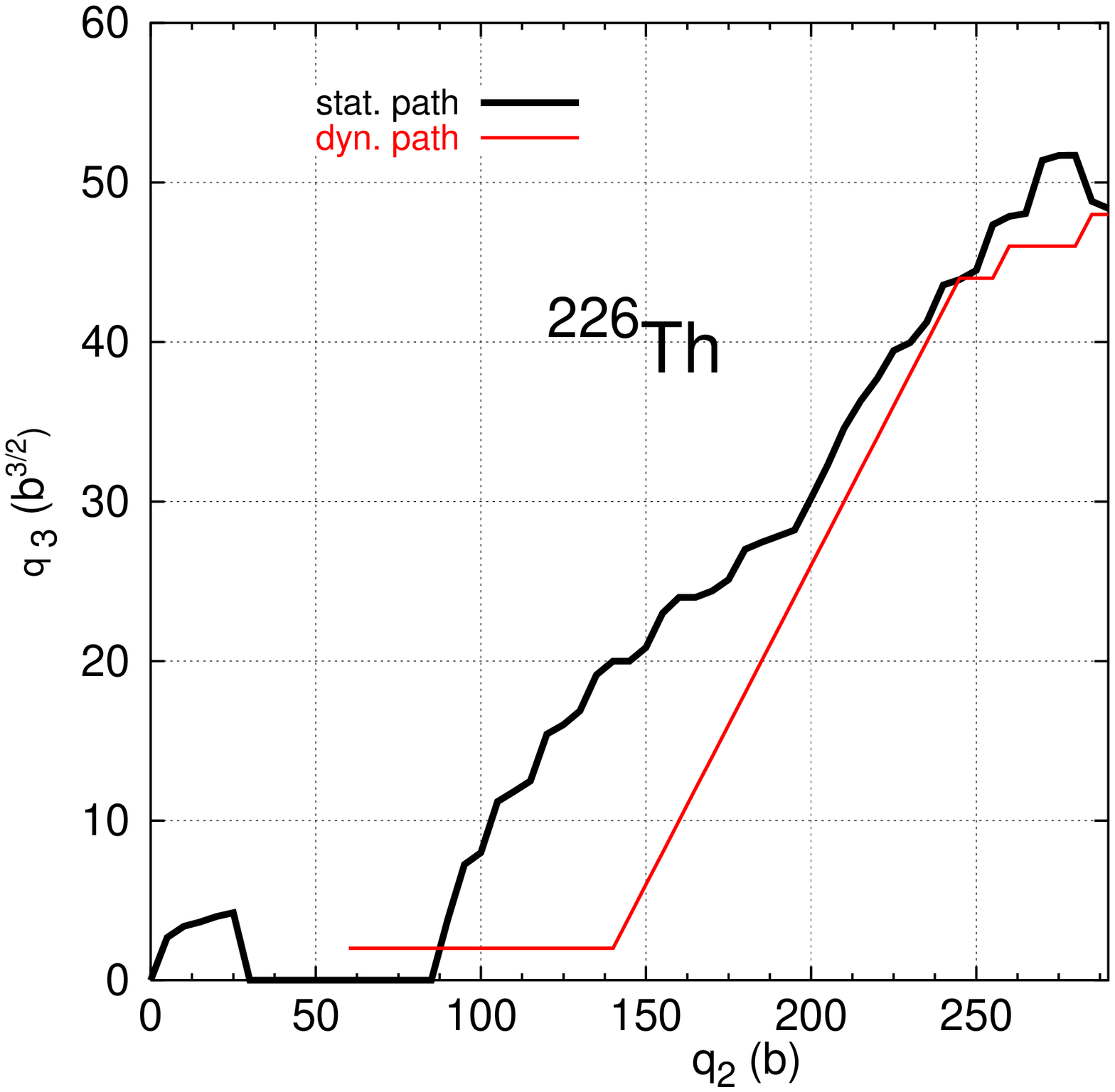, angle=270,width=8cm}}
\end{minipage}%
\begin{minipage}[c]{0.5\textwidth}
\centerline{\psfig{file=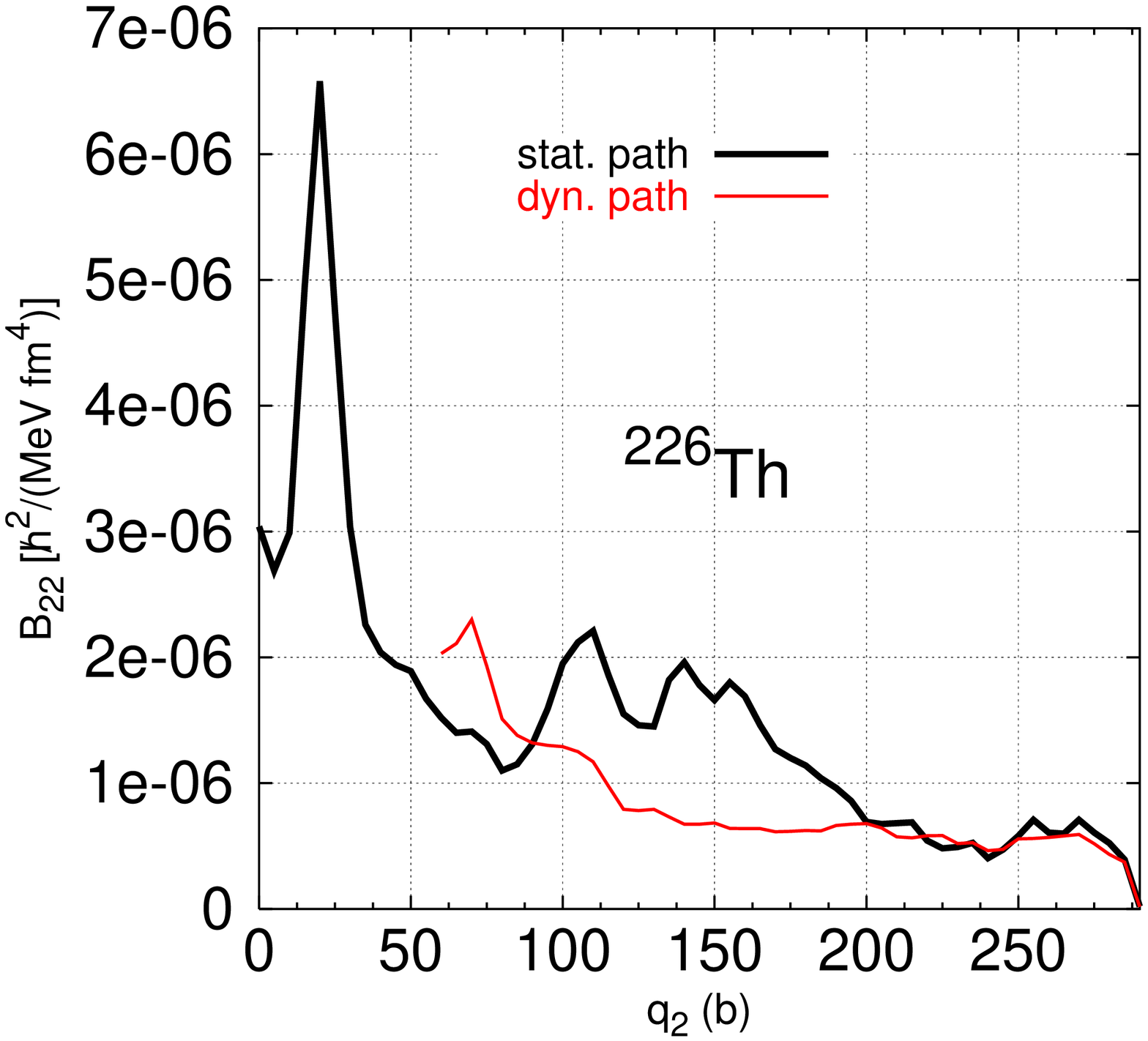, angle=270,width=8cm}}
%\hspace{-2.7cm}\includegraphics[width=3.cm, angle=270]{masa226.eps}
\end{minipage}
\vspace*{8pt}
\caption {(L.h.s. panel): the least-action trajectory $L_{min}$ (see text above) marked with thin line and the least-energy 
trajectory (thick line) obtained by minimizing the collective-potential energy $V$ of eq.~(\ref{coll-pot}) over the octupole 
deformation $q_3$, (r.h.s. panel):
leading component of the mass tensor $B_{22}$ along $L_{min}$ trajectory (thin) and along the least-energy trajectory 
(thick) for $^{226}$Th.}
\label{masy}\end{figure} 

We performed the minimization of the action integral~(\ref{action}) for two actinide nuclei, $^{226}$Th and $^{238}$U.
As mentioned before, the resulting fission path is an interplay of the behaviors of the potential energy as well as 
the collective inertia.
It is therefore difficult, at first sight, to judge where the most probable nuclear path to fission passes through 
in the energy landscape studying only its local properties.  
However, for both $^{226}$Th and $^{238}$U the symmetric and asymmetric energy valleys are relatively well pronounced 
(see Figs.~\ref{maps1}, \ref{maps2}) the behavior of their least-action fission paths (LAPs) are remarkably different (see
Fig.~\ref{bar-dyna}). 
For the $^{226}$Th isotope presented on the left hand side of Fig.~\ref{bar-dyna}, one observes that the LAP favors the 
configurations of substantially higher potential energies characterized by an about twice as much smaller inertia 
parameter (see r.h.s. panel of Fig.~\ref{masy}) rather than the least-energy states. Certainly, in this nucleus the
inertia effect plays a predominant role.
In contrast, in the right part of Fig.~\ref{bar-dyna} we show a similar study for $^{238}$U. The resulting 
LAP is now almost identical to the least-energy path. Obviously, for the latter case the fission valley is deeper and  
{\it stiffer} as compared to the analogous asymmetric valley in $^{226}$Th. Hence, the static property of 
the potential-energy landscape, undoubtedly, is of huger importance for this system, suppressing the influence 
of the collective inertia.   

The dynamic approach to the fission path can also be decisive for the problem of transitions between various energy valleys
separated from each other by a relatively low energy barrier and corresponding to different fission modes. 
In fact, for the two actinide nuclei one observes only two possible valleys corresponding to the purely symmetric and 
asymmetric fission. Nevertheless, such a transition suggested for $^{226}$Th by the static approach is finally excluded 
in the above sketched more realistic dynamical model. 
%  
%%%%%%%%%%%%%%%%%%%%%%%%%%%%%%%%%%%%%%%%%%%%%%%%%%%%%%%%%%%%%%%%%%%%%%%%%%%
%                  TRANSVERSE  MODE
%%%%%%%%%%%%%%%%%%%%%%%%%%%%%%%%%%%%%%%%%%%%%%%%%%%%%%%%%%%%%%%%%%%%%%%%%%%

\section{Collective vibrations across the fission path}

As mentioned in the introduction the nuclear fission as a collective mode
can be strongly affected by a vibrational collective movement in the direction perpendicular
to the fission path. 

The determination of the transverse mode corresponding to the symmetric fission path consists in
diagonalizing the one-dimensional collective Hamiltonian reduced from the two-dimensional one, as expressed by 
eq.~(\ref{hcoll}), by putting $i=j=3$. 
For the majority of known actinide nuclei, however, a strong mass asymmetry of fission fragments is observed.
The same treatment for an asymmetric path requires the knowledge of the local curvature of the path.
On the other hand, since the constrained HFB calculations can be only performed on the discrete grid, the local
curvature of the static fission path and the value of the inertia along the transverse path depend, to some extend, 
on the particular way of smoothing out the discrete energy values and the chosen mesh-point density.
In order to wash out the discrete structure of the energy landscape as well as the three independent components of the 
mass tensor we use the Gauss-Hermite folding method described in Ref.~\cite{approxGH}. 
The necessity of applying an interpolation, naturally leads to an uncertainty of our estimates of the transverse-vibrations 
properties. 

Having determined the set of points corresponding to the least-energy fission path by
minimizing the potential energy for a given parameter $q_2$ with respect to $q_3$ 
within the asymmetric fission valley one determines, in turn, the transversal direction $x$ which for any point on the
fission path (given by $\{q_2^{0},q_3^{0}\}$) can be locally expressed either as function of $q_2$ or as function of
$q_3$ by the simple variable transformation
\begin{equation}
x      =\left\{   \begin{array}{l}
      (q_3-q_3^{0})\frac{1}{\cos{\varphi}}, \\
	${}$                            \\ 
      (q_2^{0}-q_2)\frac{1}{\sin{\varphi}},  
      \label{q2x}\end{array}
                                         \right.
\end{equation}
where $\varphi$ is the angle between the
$q_3-$axis and the $x-$direction. Knowing the derivative of the fission path given as $q_3^0=f(q_2)$
for given $q_2^0$, we easily find the $\varphi-$values.

The approach introduced here allows, by the particular choice of the two physically relevant {\it orthogonal} collective 
degrees of freedom associated with the mode along the fission path and the transverse vibrations, 
for a more realistic prediction of the shape of the one-dimensional effective fission barriers which can serve as an input 
to the calculations of fission half-lives within the traditional WKB method. As a consequence of the separation of the 
mentioned two modes, the reduction of the full two-dimensional vibrational Hamiltonian (\ref{hcoll}) to the one-dimensional 
one of the transverse variable $x$ and the derivative $\frac{\partial}{\partial x}$ is possible with the help of the
transformation (\ref{q2x}).
In order to describe the vibrational excitations associated with the transverse collective movement,
the reduced one-dimensional (GCM+GOA) collective Hamiltonian reads
\begin{equation}
\hat{\mathcal H'}_{coll}\bigg(x,\frac{\partial}{\partial x}\bigg)=-\frac{\hbar^2}{2}\bigg( \frac{\partial}{\partial x}
\,\frac{1}{{B}_{x}}\,\frac{\partial}{\partial x}\bigg) + V_{{\rm HFB}}(x)-\Delta V(x),
\label{hcollprim}\end{equation} 
where now $\frac{\partial}{\partial x}=\cos(\varphi)\frac{\partial}{\partial q_3}$, 
$B_x$ is the $x-$component of the mass tensor which obeys the transformation
rule given already in eq.~(\ref{mass-trans}) by replacing the curvilinear coordinate $s$ along the barrier 
by the the transversal coordinate $x$
\begin{equation}
B_{x}=\sin^2(\varphi)\,B_{22}-\sin(2\varphi)\,B_{23}+\cos^2(\varphi)\,B_{33}.
\end{equation} 
The quantities $V_{{\rm HFB}}(x)$, $\Delta V(x)$ are, as in eq.~(\ref{hcoll}), the HFB potential and the ZPE corrections 
taken along the $x-$direction.
\begin{figure}[th]
\centerline{\psfig{file=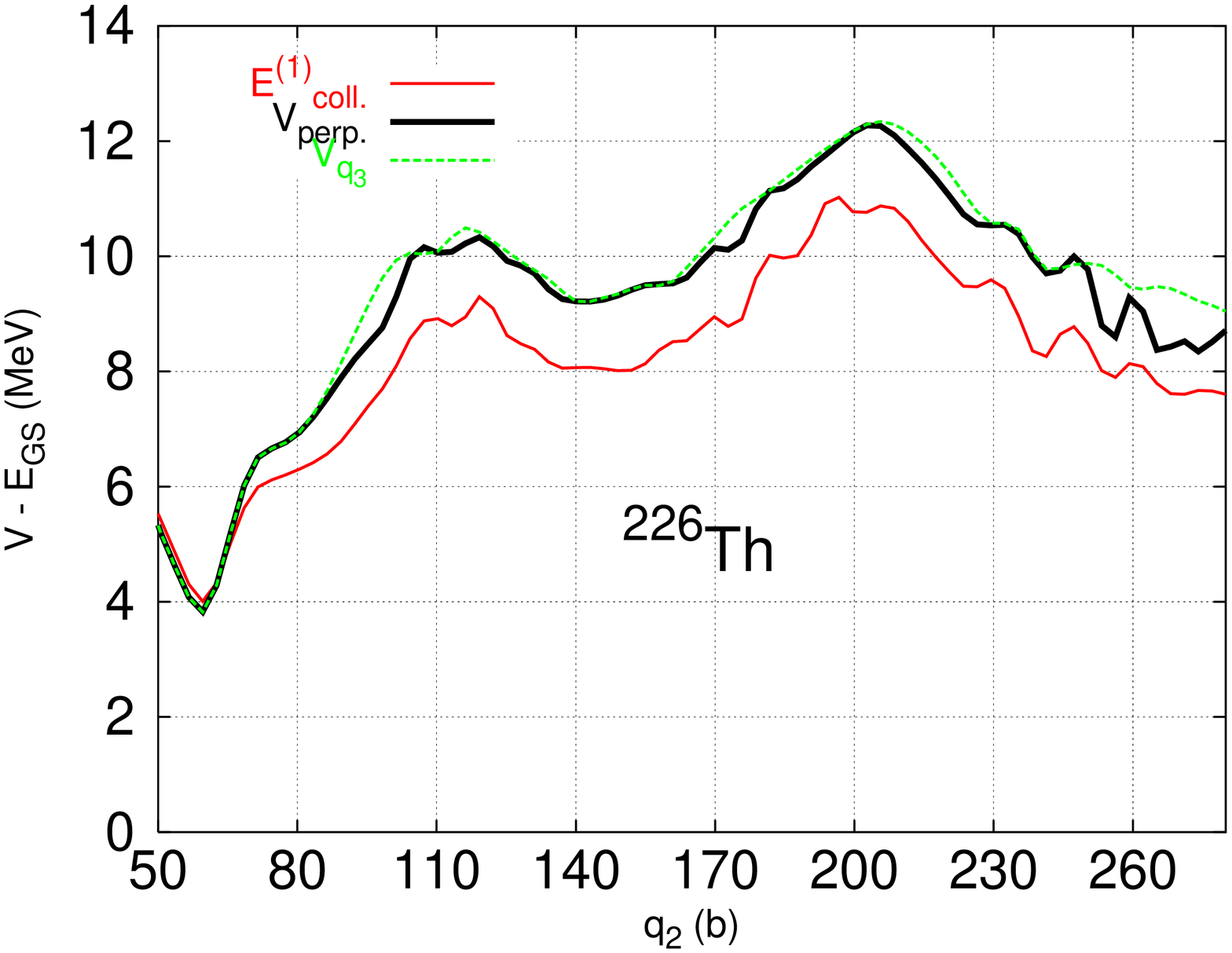, angle=270,width=8.cm}}
%\hspace{-5mm}\includegraphics[width=4.6cm, angle=270]{bars226.eps}
\centerline{\psfig{file=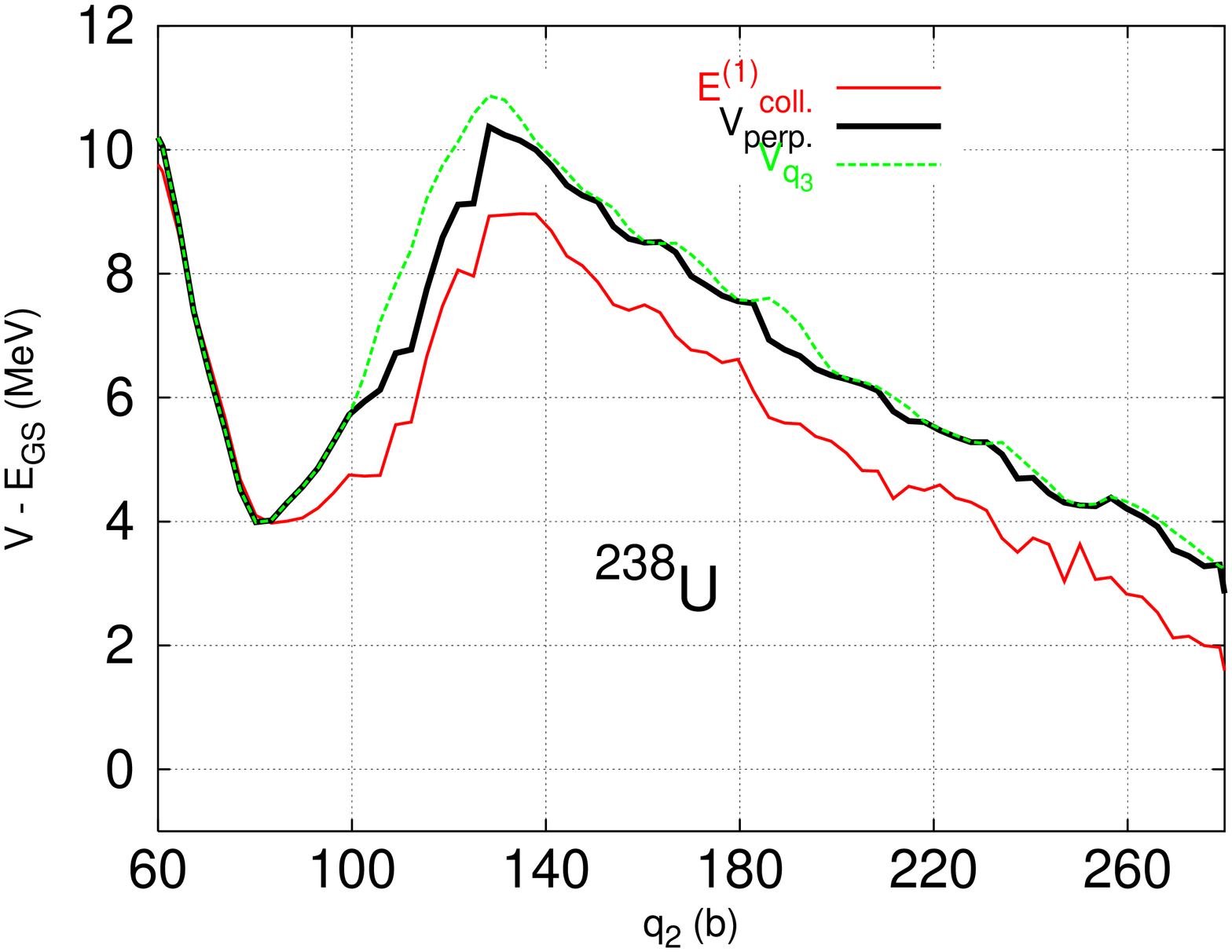, angle=270,width=8.cm}}
%\hspace{-2mm}\includegraphics[width=4.5cm, angle=270]{bars238.eps}
%\vspace*{8pt}
\caption{Static fission barrier obtained by the minimization of the potential energy given by 
eq.~(\ref{coll-pot}) with respect to $q_3$ parameter (dashed line),
static barrier as a result of the potential-energy minimization along the perpendicular to the static 
fission path direction $x$ for given $q_2-$value (thick full line), the barrier with transverse vibrations 
(thin full line)}
\label{Fig-3}\end{figure}

Figure \ref{Fig-3} illustrates three fission barriers (all artificially shifted in energy to coincide at the isomeric minima) 
obtained in different manners. The barrier marked with thin dashed line
is obtained by minimizing the collective-potential energy $V_{{\rm HFB}}(q_2,q_3)+\Delta V(q_2,q_3)$ 
(with $\Delta V$ denoting the sum of the vibrational and rotational zero-point corrections) with respect to the 
octupole deformation $q_3$ while the 
thick full line corresponds to the minimum of this energy as a function of $q_2$ measured along the direction perpendicular to 
the fission path. The third barrier represents the energy of the lowest vibrational states depending 
on $q_2$ and resulting from the diagonalization of the one-dimensional collective Hamiltonian (\ref{hcollprim})
across the fission path.
As one could expect, the last mentioned barrier including the effect of transverse vibrations is effectively the 
lowest. This becomes clear when one realizes that the density of low-lying collective states 
built in the isomeric well is smaller as compared to the density in the saddle configuration. 
This directly implies that the distance between the collective potential energy minimum and the first vibrational 
state of the transverse mode in the isomeric well is larger as compared to the corresponding distance for the 
outer-saddle configuration. 

Notice that the above mentioned effect can be easily explained in terms of the harmonic approximation where the 
energy interval between subsequent vibrational states is proportional to the vibrational frequency
depending on the {\it stiffness} of the potential along the transverse direction in the vicinity of the isomeric 
minimum and the inverse of the effective mass parameter in this region. 
Therefore, the effective barrier height, measured as the difference between the lowest vibrational states of the isomeric 
minimum and the saddle configuration is, in fact, lower than for the static barriers shown in Fig.~\ref{Fig-3} 
by the order of 1 MeV for both investigated nuclei.

\section{Summary and conclusions}

We have presented three approaches to determine the effective fission path, demonstrating in this way 
a possible improvement of the commonly applied one-dimensional static approach to fission barriers. 

We have in particular confirmed in section \ref{LAP} that, due to the balance between
the static properties of the potential landscape and the collective inertia, significant changes can take place
in the way nuclei to fission, particularly for nuclei characterized by relatively weakly pronounced energy valleys 
which determine the static fission paths on the average.
The example of the nucleus $^{226}$Th has shown that in that case the least-action path leads through the states characterized by an
about 3-5 MeV higher potential energy than the least-energy path. 
Keeping in mind the one-dimensional WKB method, one easily concludes that the collective inertia is of the same importance 
for the barrier penetrability problem as the shape of the fission barrier. 

We have observed for the case of $^{226}$Th that a purely static treatment of the fission phenomenon would allow 
for transitions between different fission valleys, separated by a substantially high energy barrier.  
In this particular case the rapid shift between different fission modes would lead to two sudden modifications of the 
mass distribution of the fission fragments due to a slight variation of the elongation.
Using the method based on the action integral presented in section \ref{LAP} one demonstrates that in $^{226}$Th 
such an unlikely transition does not exist. 
Nevertheless, according to our recent preliminary investigations, similar single transitions between asymmetric and symmetric
modes might exist in the region of Fm isotopes. Investigations devoted to this topic are under progress.

As already mentioned in the introduction, the one-dimensional description of the fission-barrier penetrability seems to 
be not sufficient since the theoretical predictions of, for example, fission half-lives in different mass regions remain
in disagreement with the experimental estimates by several orders of magnitude. 
Once the fission path is found we attempt in section 5 to extend this approach by considering the influence of the collective 
transverse vibrations along this path using fully microscopic GCM+GOA approximation well suited for studies of heavy and 
super-heavy nuclei. 
Such an approach provides the consistent form of the collective vibrational Hamiltonian without any adjustable parameters.  
As expected, the collective transverse movements cause a lowering of the effective one-dimensional fission barrier 
height by approximately 1 MeV for the discussed $^{226}$Th and $^{238}$U nuclei. This happens due to different static 
(potential energy stiffness) and dynamical 
(inertia) collective properties of the isomeric minimum and the saddle point of the fission valley.
A collective effect of that order of magnitude can by no means be neglected in the estimates of 
fission half-lives or fission cross sections.

\end{document}